\documentclass[nofootinbib,twocolumn,prd]{revtex4}
\usepackage{hyperref}
\usepackage{amsmath}
\usepackage{graphicx}
\usepackage{color}
\usepackage{ifpdf}

\newcommand\editremark[1]{ {\color{red} #1}}

\newcommand\optional[1]{}
\newcommand\ForInternalReference[1]{}

\newcommand\unit[1]{\, {\rm #1}}

\newcommand\Y[1]{Y^{(#1)}}

\newcommand\avL{\left< L_{(a} L_{b)} \right>}
\newcommand\WeylScalar{{\psi_4}}

\newcommand\qmstate[1]{\left|#1\right>}
\newcommand\qmstateKet[1]{\left<#1\right|}
\newcommand\qmstateproduct[2]{\left<#1|#2\right>}
\newcommand\qmoperatorelement[3]{\left<#1\left|#2\right|#3\right>}

\newcommand{\ROS}[1]{#1}

\def\bbh#1{binary black hole#1 (BBH#1)\gdef\bbh{BBH}}
\def\bh#1{black hole#1 (BH#1)\gdef\bh{BH}}

\begin{document}
\title{Efficient asymptotic frame selection for binary black hole spacetimes using asymptotic radiation} 
\author{R.\ O'Shaughnessy}
\affiliation{Center for Gravitation and Cosmology, University of Wisconsin-Milwaukee,
Milwaukee, WI 53211, USA}
\email{oshaughn@gravity.phys.uwm.edu}
\author{B. Vaishnav}
\altaffiliation{Current address: School of Physics, Georgia Southern University, Statesboro, GA 30458, USA}
\author{ J. Healy}
\author{ Z. Meeks}
\author{D. Shoemaker}
\affiliation{Center for Relativistic Astrophysics,
Georgia Tech, Atlanta, GA 30332, USA}

\begin{abstract}
Previous studies have demonstrated that gravitational radiation reliably encodes information about the natural emission
direction of the source (e.g., the orbital plane).   In this paper, we demonstrate that these orientations can be
efficiently estimated by  the principal axes of $\avL$, an average of the action of rotation group generators on the
Weyl tensor at asymptotic infinity.    Evaluating this average at each time provides the
instantaneous emission direction.   Further averaging across the entire signal yields an average
orientation, closely connected to the angular components of the Fisher matrix.  The latter direction is well-suited to data
analysis and parameter estimation when the instantaneous emission direction evolves significantly.
Finally, in the time domain, the average $\avL$ provides fast, invariant diagnostics of waveform quality.
\end{abstract}
\keywords{}

\maketitle

\optional{
Citations to consider:

 - spinning particles EOB: \cite{2010PhRvD..81h4024B}

 - belt of real numbers
}

\section{Introduction}
Ground-based gravitational wave detectors like LIGO and Virgo are  likely to see many  few-stellar-mass black hole (BH) binaries
formed through isolated \cite{PSellipticals,popsyn-LowMetallicityImpact-Chris2010} and dynamical
\cite{clusters-2005,2008ApJ...676.1162S,2010MNRAS.402..371B} processes.  Additionally, advanced detectors could see the merger signature of
two intermediate-mass black holes (each $M\in[100, 10^3] M_\odot$), perhaps  formed in dense globular
clusters \cite{imbhlisa-2006}.  
Black hole spin will significantly impact all phases of the signal to which these ground-based detectors are sensitive:  the
late-time inspiral, merger signal, and (through the final BH spin) ringdown. 
Misaligned spins break symmetry in the orbital plane, generally leading to strong modulations
in the inspiral \cite{ACST}, merger, and ringdown.  \optional{others}  
At early times, these modulations can be easily estimated through separation of timescales: the emitted inspiral waveforms are
often well-described by quasistationary radiation from a circular orbit, slowly rotating with time modulated with time
as the orbital plane precesses
\cite{gwastro-mergers-nr-ComovingFrameExpansionSchmidt2010,ACST,gw-astro-mergers-approximations-SpinningPNHigherHarmonics}.
In the extreme but astrophysically important case of BH-NS binaries, the emission direction can precess about the total
angular momentum through arbitrarily large angles, leading to strong amplitude and phase modulations \cite{ACST,gw-astro-SpinAlignLundgren-2010}.
At late times,  numerical relativity simulations of mergers can efficiently predict gravitational radiation from a
variety of spins and orbits.   A well-chosen static or time-evolving frame makes it easier to compare their outputs to
one another and to these analytic models. 
Previous studies  have demonstrated the gravitational radiation carried to infinity  encodes information about
the instantaneous emission directions.  For example,  
\citet{gwastro-mergers-nr-ComovingFrameExpansionSchmidt2010}  track the precession of the orbital plane by selecting a
frame that maximizes the amplitude of the $l=|m|=2$ modes at each time.    In this paper, we propose an  invariant algebraic
method to calculate the natural principal axes of asymptotic radiation.  This method  can be applied  independently at each time,
at each frequency, or at each  binary mass.
 As with previous studies \cite{gwastro-mergers-nr-ComovingFrameExpansionSchmidt2010} the diagonalized time-domain representation 
may be useful in interpreting and modeling precessing signals as quasistationary circular orbits.
Data analysts, however, will benefit most from choosing an optimal frame at each binary mass.   For example, some signals like BH-NS
binaries have an optimal time-domain emission orientation that evolves significantly.  In these cases, as well
as when higher harmonics become significant at high mass, the frame relevant to \emph{data analysis} must suitably
average over the relevant portion, frequency, and modal content of the signal.   In this paper, we provide the first
algorithm to select a suitable orientation for  the average signal.   

In Section \ref{sec:Method}  we describe our algebraic method for determining an optimal orientation versus
time, frequency, or mass.
In Section \ref{sec:Time} we demonstrate  our method provides a new way to efficiently and accurately determine the natural orientation of a
quasistationary quasicircular inspiral versus time, using both artificially generated quasistationary sources (i.e.,
applying a rotation to a nonprecessing source) and waveforms from real precessing binaries.
We also discuss  numerical limitations in our time-domain method during the merger and ringdown, where multiple harmonics become increasingly important.
In Section \ref{sec:Mass} we demonstrate that a preferred frame versus \emph{mass} arises naturally in data analysis
applications, as the average of the Fisher matrix.   Using a separation of timescales, we demonstrate this frame roughly corresponds to a
time- and bandpass-weighted average of the optimal time-dependent frame. 
We briefly comment on the utility of mass-weighted frames for data analysis applications, such as mode-decomposed
searches for spinning binaries \cite{BCV:PTF}.

\section{Natural waveform frames}
\label{sec:Method}
Gravitational radiation carried away to infinity is well-known to encode certain preferred orientations, such as the
radiated linear and angular momenta.  More generally, given a one-parameter family of inner products
$\qmstateproduct{a}{b}_\xi =\int a^*(t')K_\xi (t,t')b(t') dt dt' $ generated
by $K_\xi(t,t')$ and a tensor operator acting on the asymptotic Weyl scalar, the orientation average of
$\qmstateproduct{\WeylScalar}{Q_{a_1\ldots a_n}\WeylScalar}$ defines an asymptotic tensor
\begin{eqnarray}
\left<{Q_{a_1\ldots a_n}}\right>_{\xi} &\equiv&  \frac{ 
\int d\Omega  \int dt dt' K_\xi(t,t')\WeylScalar^* Q_{a_1\ldots a_n}\WeylScalar
 }{
 \int d\Omega  \int dt dt' K_\xi(t,t')\WeylScalar^* \WeylScalar
} \; .
\end{eqnarray}
The principal axes of this tensor in turn define preferred directions.   
Depending on the inner product and reference tensor, very different orientations can be selected; this formula, for
example, includes as a special case both recoil kicks and radiated angular momentum.

In this paper we consider $\avL$, the average of products of \emph{rotation group generators} $L_k$.\footnote{Explicit coordinate forms 
for the generators $L_k$ acting on functions with spin weight $s$ are provided by \cite{gr-nr-Multipoles-Alcubierre2007},
denoted as $J_k$.
As the algebra of generators acting on the representation $\Y{s}_{lm}$ is unchanged, only the usual action of $L_k$ on
rotation eigenstates is needed to perform the calculations described in this paper.}
For example, adopting a kernel $K_t(\tau_1,\tau_2)=\delta(\tau_1-t)\delta(\tau_2-t)$ (i.e., using a
quantum-mechanics-motivated state notation,  $K =
\qmstate{t}\qmstateKet{t}$), an average orientation at each time $t$ can be calcualted from the principal axes of
\begin{subequations}
\begin{eqnarray}
\left< L_{(a} L_{b)} \right>_t &=& 
 \frac{\int d\Omega \WeylScalar^*(t) L_{(a}L_{b)} \WeylScalar(t)
  }{
   \int d\Omega |\WeylScalar|^2
}
 \\
 &=& \frac{\sum_{lmm'} \WeylScalar_{lm'}^*  \WeylScalar_{lm}\qmoperatorelement{lm'}{L_{(a}L_{b)}}{lm} }{\int d\Omega |\WeylScalar|^2}  \nonumber
\end{eqnarray}
where in the second line we expand $\WeylScalar= \sum_{lm} \WeylScalar_{lm}(t)\Y{-2}_{lm}(\theta,\phi)$ and perform the angular integral.  In this
case, the average tensor $\left< L_{(a} L_{b)} \right>_t $ at each time $t$ can be calculated algebraically, from the
mode amplitudes $\WeylScalar_{lm}$ and  the well-known action of SU2 generators $L_k$  on rotational eigenstates.  The
principal axes follow by diagonalizing this $3\times 3$ matrix.  
The same construct can be applied to any kernel $K$.   In particular, we define the average of $L_a L_b$ at any
frequency $f$ or over all frequencies by
\begin{eqnarray}
\left< L_{(a} L_{b)} \right>_f &=& 
 \frac{\int d\Omega \tilde{\WeylScalar}^*(f) L_{(a}L_{b)} \tilde{\WeylScalar}(f)
  }{
   \int d\Omega  |\tilde{\WeylScalar}(f)|^2
}\\
\left< L_{(a} L_{b)} \right>_M &=& 
 \frac{\int d\Omega \int df \frac{\tilde{\WeylScalar}^*(f) L_{(a}L_{b)} \tilde{\WeylScalar}(f)} {(2\pi f)^4 S_h(f)}
  }{
   \int d\Omega  \int df |\tilde{\WeylScalar}|^2/ [(2\pi f)^4S_h(f)]
}
\end{eqnarray}
\end{subequations}
where $S_h$ is the strain noise power spectral density of a gravitational wave detector.  
\ROS{
The subscript $M$ in the second expression denotes the binary mass, used to connect the scale-free output
 of a numerical
relativity simulation to physical time $t$  (i.e., $r M \WeylScalar(t)$ depends only on $t/M$).
For context, in gravitational wave data analysis, the inner product $\qmstateproduct{a}{b}=2 \int a^*(f) b(f)/S_h(f)$
characterizes the natural optimal-filtering-induced metric for a noise power spectrum $S_h$;  the denominator in the
final expression thus corresponds to the inner product $\qmstateproduct{h_+}{h_+}+\qmstateproduct{h_\times}{h_\times}$
where the strain  $\int \int
\WeylScalar dt dt' \propto h_+ - i h_\times$ is reconstructed from $\WeylScalar$ by integration.

}

The components, eigenspaces, and eigenvalues of this tensor  characterize the emitted radiation, averaged over orientation and the
specified quantity $\xi$ ($=t,f,M,\ldots$).   As  components of this matrix are associated with averages of mode
orders weighted by $|\WeylScalar|^2$, such as 
\begin{eqnarray}
\left<L_z^2\right> &=& \left<m^2\right> \\
\text{Tr}\left<L_a L_b \right> &=& \left< l(l+1)\right>
\end{eqnarray}
the components of $\avL$  identify both the dominant emission direction and the dominant emission mode.
In the special case of  quasistationary emission along an $l=|m|=2$ mode aligned with $\hat{z}$, the  eigenspaces of $\left<
L_{(a} L_{b)} \right>_t$ are are $\hat{z}$ (eigenvalue $m^2=4$) and the $x-y$ plane (eigenvalue
$\left<l(l+1)-m^2\right>/2=1$).    
As quadrupolar emission dominates, these statements are a good approximation for all time-domain averages $\avL_t$.  

  Small errors in high-order modes can be amplified by the factor
$l^2$, particularly during the merger phase where higher harmonics become significant both physically and through
  numerical error.  However,
since higher-order $l$ subspaces can be invariantly removed from  $\WeylScalar$, a similar average can also be constructed
for any subset of $l$.   To the extent that the dominant emission orientation is well-resolved, the directions derived from
restricted averages should agree.

\section{Optimal orientation versus time }
\label{sec:Time}

\begin{figure}
\includegraphics[width=\columnwidth]{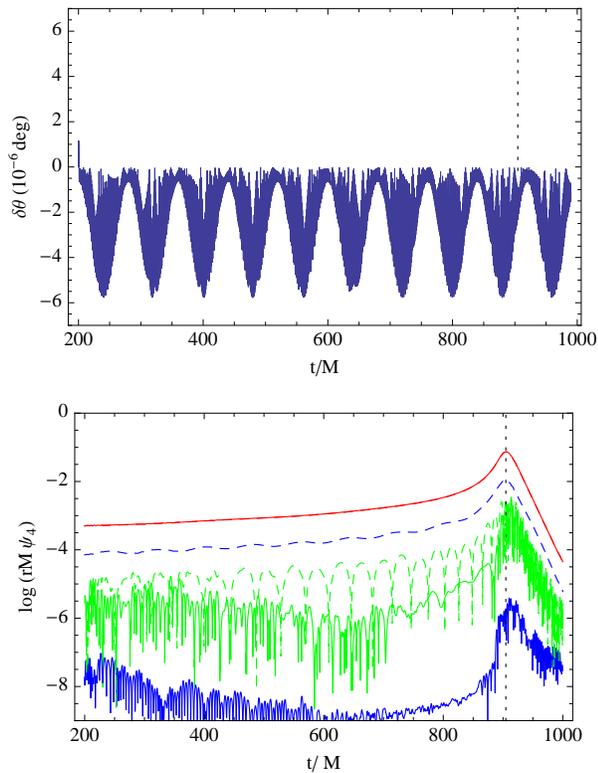}
\caption{\label{fig:Time:Synthetic}\textbf{Recovering a synthetic rotation}: Starting with a equal-mass
nonspinning ($a=0.0$)  binary waveform (bottom panel: $\WeylScalar_{lm}$ for $\ell \le 2$, extracted at $r=60 M$), we apply a rotation
taking $\hat{z}$ to the polar angles $(\theta,\phi)$, where $\theta(t)=\theta_o(1+0.1 \cos 2\pi t/P)$  and $\phi(t)=2\pi t/P$
for $P=80 M$ and $\theta_o=\pi/20$.   The bottom panel shows $\WeylScalar_{lm}$ after (dashed) and before (solid) this rotation, for the modes
$l=2$ and $m=2$ (red), $1$ (blue), and 0 (green).  Using $\avL_t$, we recover an orientation $\hat{Z}(t)$ from
the rotated waveform that agrees with the imposed rotation to within $5\times 10^{-6}\unit{deg}$  (top panel:
$\theta(Z(t))-\theta(t)$).  Applying this rotation recovers the unperturbed
waveform. 
A vertical dotted line indicates the peak $l=|m|=2$ emission in both figures.
}
\end{figure}

\begin{figure}
\includegraphics[width=\columnwidth]{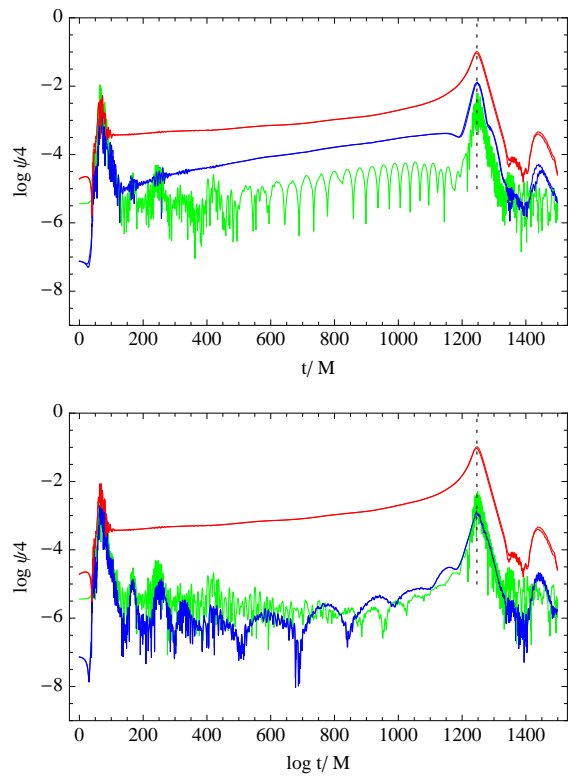}
\caption{\label{fig:Time:Real} \textbf{Aligning a precessing binary}: Starting with a precessing equal mass binary
  ($a_1=0.6 \hat{z},a_2=0.6(\hat{x}+\hat{z})/\sqrt{2}$; top panel), we derive the principal axis orientation from $\avL_t$,
  restricted to $l=2$ modes.  The bottom panel shows the transformed harmonics.   The orientation transformation is derived and applied at all
  times shown.   \ROS{To illustrate this method's robustness, in this figure we retain initial transients and late-time errors.}
}
\end{figure}

The dominant principal axis of $\avL_t$ invariantly determines a preferred orientation $\hat{\cal Z}(t)$ versus time.
Moreover,  by design, any waveform that is well-approximated by quasistationary emission of a  $l=|m|=2$ mode  along
$\hat{Z}(t)$ must have $\hat{\cal Z}(t)$ correspond to that instantaneous emission symmetry direction.
On the contrary, though a priori the eigensystem of the symmetric tensor $\avL$ provides a frame, empirically the  directions associated with the two smaller eigenvalues cannot be trusted.   During the inspiral the perpendicular directions are degenerate; small numerical
asymmetries in each timestep cause the two smaller eigenvectors to jump chaotically in the plane perpendicular to the
dominant eigenvector.
Given the empirical degeneracy of the smaller two eigenvalues, we adopt a preferred frame at time $t$ by a fiducial rotation:
starting with a cartesian frame aligned with the initial simulation's coordinate frame,  rotate  by $R =
R_z(\phi)R_y(\theta)$, which takes $\hat{z}$ to $ \hat{Z}$ if $\theta,\phi$ are the spherical polar coordinates for $\hat{Z}$.
\ROS{In terms of this rotation, the waveform in the instantaneous ``aligned'' frame at each time can be expressed in
  terms of the instantaneously ``aligned'' eigenstates $\Y{-2}_{LM}(R^{-1} n)$  and the simulation-frame expansion
  coefficients $\WeylScalar_{LM}$ using the representation theory of SU(2) \cite{gr-nr-RotatingGWEmission-Berti}:
\begin{eqnarray}
\WeylScalar(t,\hat{n}) &=& \sum_{LM} \WeylScalar'_{LM} \Y{-2}_{LM}(R^{-1}\hat{n}) \\
\WeylScalar_{LM}'& =& e^{2 i \chi(R)}\sum_{M'}   D^L_{MM'}(R)  \WeylScalar_{LM'}
\end{eqnarray}
The overall phase $\chi$ does not enter into any calculation we perform and will be ignored.
}

As a concrete example, in Figure \ref{fig:Time:Synthetic} we compare the difference between a generic time-dependent rotation and
the reconstructed orientation, obtained by applying that rotation to a nonprecessing black hole binary simulation and
reconstructing the optimal direction versus time.    The quasicircular orientation is recovered to $\delta \theta \simeq
10^{-6}\unit{deg}$ (i.e., $\delta \theta^2$ comparable to working machine precision).
As another example, in Figure \ref{fig:Time:Real}, we compare the mode amplitudes for a spinning, precessing binary in an asymptotic
inertial frame (top) and in the ``corotating'' frame implied by this transformation (bottom). %
The transformed modes  resemble the modes of a spin-aligned binary (Fig   \ref{fig:Time:Synthetic}, right
panel). 
Note that in going from the generic-spin to ``corotating'' frame, the $m=1$ and $m=0$ modes
  have been reduced by more than an order of magnitude during the inspiral.

 \citet{gwastro-mergers-nr-ComovingFrameExpansionSchmidt2010} have previously reconstructed optimal orientations from the
 emitted waveform.  By contrast to their maximization-based method for tuning $\WeylScalar$, our algebraic method is fast,
 accurate, and invariant.  
This method can also be applied to any constant-$l$ subspace.  For example, using the GT/PSU equal-mass, generic spin simulation set summarized
by \citet{gwastro-spins-rangefit2010} we have reconstructed $\avL$ for all $l\le 4$ modes; for  $l=2$ only; and for $l=4$
only.\footnote{We have also tested the behavior of $\avL_t$ using only the $l=3$ mode subspace.  During inspiral, this
  tensor has a well-defined dominant principal axis that agrees with the $l=2$ result.  However, at and beyond merger,
  $\avL_t$ becomes nearly isotropic and does not provide a natural emission direction.}   All agree during the inspiral and at least some of the merger.\footnote{A few tens of $M$ after the merger, owing to their low and
  rapidly decaying  amplitude, the $l=4$ modes previously available are not resolved to the precision needed for
  directional reconstruction.  }

Finally, as this method operates on each $l$ subspace as well as independently and algebraically at each timestep, it
provides valuable, fast, invariant diagnostics of the quality of numerical waveforms.    For example, numerical
simulations can evaluate their waveform quality by the following three constructs: (a) eigenvalues of
$\avL_t$ should evolve smoothly and remain  near their predicted values (i.e., near $4,1,1$ if the $l=2$ subspace is included); (b) the recovered dominant
eigendirection of $\avL_t$   should be smooth in time; and (c) phenomenologically speaking, the recovered paths from different multipole orders should agree.

\section{Optimal orientation versus mass }
\label{sec:Mass}

In many cases of astrophysical interest, the emission direction changes significantly during the merger.  As an example,
BH-NS binaries can have their dominant emission direction, the orbital angular momentum direction $\hat{L}$, precess one to
several times in a cone around the total angular momentum \cite{gw-astro-SpinAlignLundgren-2010}.    In this case in
particular and for high mass ratio binaries in general, the instantaneous emission direction averages out.  Instead, the natural
direction encoded by the signal  is the total angular momentum.
Generally, however, a suitable average is not intuitively self-evident.

A detector with comparable and coherent sensitivity to both gravitational wave polarizations is naturally characterized
with a complex inner product that coherently accounts for both polarizations \cite{gwastro-mergers-nr-CoveringSpace-ROS-Methods}.
Specifically, two gravitational waveforms $A,B$ expressed as curvature ($\WeylScalar$) along a particular line of sight $\hat{n}$ are naturally compared by a
complex inner product 
\begin{eqnarray}
(A,B) &\equiv&  \int_{-\infty}^{\infty} 2 \frac{df}{(2\pi f)^4 S_h} \tilde{A}(f)^* \tilde{B}(f)
\end{eqnarray}
Taking derivatives with respect to angle and expanding out the generators of the rotation matrix, the Fisher matrix
($\Gamma_{ab} = |\left(\partial_a\WeylScalar,\partial_b\WeylScalar\right)|/|(\WeylScalar,\WeylScalar)|$) reduces to 
\begin{eqnarray}
\Gamma_{ab} = \frac{|( L_{(a} \WeylScalar,L_{b)}\WeylScalar)|}{|(\WeylScalar,\WeylScalar)|} 
\end{eqnarray}
The  tensor  $\avL_M$ therefore corresponds to the orientation-averaged Fisher matrix.\footnote{For brevity, we do not
  discuss maximization over coalescence time and phase; see \cite{gwastro-mergers-nr-CoveringSpace-ROS-Methods} for a
  more extensive discussion of the Fisher matrix derived from a complex overlap.}   In other words, $\avL_M$ characterizes  how well we can determine a binary's orientation, averaged over
all possible directions along which we could see it.

Moreover, for quasistationary precession dominated by a single mode at frequency $f(t)$, the signal-weighted  average $\avL_M$ has a natural physical interpretation as
a power-weighted average of $\avL_t$.   Substituting the stationary phase approximation to the Fourier transform 
\begin{eqnarray}
\tilde{\WeylScalar}(f)&=& \frac{1}{\sqrt{i df/dt}}\WeylScalar( t(f))  e^{-i 2\pi f t(f)}
\end{eqnarray}
into the overlap $\left(L_a\WeylScalar,L_b\WeylScalar \right)$ leads to an amplitude-weighted integral over $\avL_t$
\begin{eqnarray}
\left(L_a\WeylScalar,L_b\WeylScalar \right) &\simeq& \int  \frac{2 dt}{(2\pi f(t))^4 S_h(f(t))}(L_a \WeylScalar(t))^* (L_b \WeylScalar(t))  
\end{eqnarray}
The orientation average of the Fisher matrix is therefore 
\begin{eqnarray}
 \int \Gamma_{ab} d\Omega/(4\pi) &\simeq& \frac{1}{|(\WeylScalar,\WeylScalar)|}
\int 2 dt \frac{\avL_t}{(2\pi f(t))^4 S_h(f(t))}
\end{eqnarray}
In other words, for sufficiently steady quasicircular precession, $\avL_M$ is precisely the power-weighted average of
$\avL_t$.  The principal eigenvector of $\avL_M$ therefore corresponds to  our intuition about an ``average frame''.

\section{Conclusions}
In terms of  orientation-averaged  tensors, we have introduced an efficient algebraic method to extract fiducial
orientations from waveforms extracted to infinity.   Specifically, we have
demonstrated that one  natural orientation of the emitted beampattern  is associated with the dominant principal axis of
$\avL$.  
The average can be calculated at any time or averaged, at any mass.  In the time domain, our method accurately,
efficiently, and smoothly reconstructs the emission direction throughout the inspiral.  As with
\cite{gwastro-mergers-nr-ComovingFrameExpansionSchmidt2010}, if we retroactively align the waveform with this optimal
direction at each timestep, the dominant modes become smoother and the subdominant modes significantly smaller.

Conversely, if applied at each mass, our method also has close, provable connections to data analysis: it corresponds to
the Fisher matrix for orientation angles, averaged over all possible  lines of sight.   In the case of extreme
mass ratio, except for rare high-symmetry nonprecessing inclined orbits, our preferred direction reduces to the total angular momentum, about which the orbit precesses.  
This invariant approach to the preferred orientation will benefit search strategies for numerical relativity
waveforms from strongly precessing binaries that coherently employ multiple modes, analagous to the proposal by \citet{BCV:PTF} for strongly precessing
BH-NS binaries.

We have proposed one of many possible definitions for the ``instantaneous emission direction.''    Of course, other
plausible generating tensors $Q$ can be constructed with $L_a$ alone, each
corresponding to a different way of weighting the mode amplitudes $\WeylScalar_{lm}$.  Still more can be added by broadening
the space of operators.   That said, our definition offers a so-far unique feature:  a provable connection to an
astrophysically relevant quantity (the Fisher matrix).  

Other astrophysically significant features of numerical relativity waveforms could be employed to determine preferred directions.  For example, a ``generalized equatorial plane'' could consist of the
set of directions along which gravitational radiation is locally linearly polarized (e.g., as the surface $\partial_t \ln \WeylScalar/\WeylScalar^*=0$).  Like the average $\avL$, these directions  can be defined in the time, frequency, and mass
domain.   These emission directions lead to parameter estimation degeneracies: equal amount of left- and right-handed radiation imply two possible binary
orientations are consistent with that signal.   Though intriguing and invariant, these symmetry directions depend sensitively on
delicate cancellations of all available modes.
Less invariant but equally significant is a direct comparison  (``overlap'') of waveforms emitted in all possible
directions, as would be seen by a gravitational wave detector.  This  method will be discussed in a subsequent
paper \cite{gwastro-mergers-nr-CoveringSpace-ROS-Methods}, in the broader context of generic waveforms from spinning binaries.

\ROS{%
Finally and as discussed previously by \citet{gwastro-mergers-nr-ComovingFrameExpansionSchmidt2010}, both radiation and
merger physics are more easily modeled in an ``aligned'' frame.  
For example, at present, hybrid waveforms have been constructed primarily for spin-aligned binaries for a handful of harmonics
\cite{gwastro-nr-Phenom-Lucia2010,gwastro-Ajith-AlignedSpinWaveforms} (cf. \citet{2010JPhCS.243a2007S}).  By tabulating and
modeling the ``aligned''-frame waveforms and the corotating frame itself, hybrids can be constructed for generic
precessing waveforms.
As another example,  previous studies of nonspinning and aligned-spin binaries suggest that their low-order modes
evolve in phase  with each other ($\arg \WeylScalar_{lm} \propto m \arg \WeylScalar_{22}$) through inspiral and
merger \cite{2008PhRvD..78d4046B,gwastro-mergers-nr-AllModesCommonPhasing-Kelly2011}.    By analogy, when expressed in a
an ``aligned'' frame, low-order modes of generic precessing binaries also seem to evolve in phase.   We will address
these and other physical properties of  ``aligned''-frame waveforms in a subsequent publication.
}

\optional{
\appendix
\section{Applying Rotations to Harmonics}
To validate our choice of basis frame, we explicitly performed rotations of $\WeylScalar$ using well-known transformation
rules of spin-weighted spherical harmonics in terms of $D$ operators.  Specifically, if $R$ a rotation transforming
the $z$ axis to $\hat{r}$ for spherical polar coordinates ($\hat{z}\rightarrow \hat{r}$), we know  $\Y{-s}_{lm}(R\hat{r})\propto
D_{-s,m}^l(R^{-1})$.  The transformation properties of the $D$ operators implies the basis functions and basis coefficients transform according to
\begin{eqnarray}
e^{i s \chi} \Y{s}_{lm}(R^{-1}\hat{n}) &=& \sum_M  \Y{s}_{lM}(\hat{n}) D^j_{Mm}(R)  \\
\WeylScalar_{lm}(R^{-1}) &=& \sum_{M}  D^l_{mM}(R) \WeylScalar_{lM}
\end{eqnarray}
The function  $\chi(R,\hat{n})$ will not enter into any computation we perform.

\editremark{VALIDATE DIRECTION SIGNS}: example of arbitrary rotation and its inverse being neede.
}

\begin{acknowledgements}
DS is supported by NSF  awards PHY-0925345, PHY- 0941417, PHY-0903973 and TG-PHY060013N.
BV would like to acknowledge the financial support of the Center for Relativistic Astrophysics.
ROS is supported by NSF award PHY-0970074, the Bradley Program Fellowship, and the UWM Research
Growth Initiative.  ROS also thanks Aspen Center for Physics, where
this work was completed, and E. Berti for helpful discussions
\end{acknowledgements}

\bibliography{paperexport}
\end{document}